\documentclass[twocolumn]{revtex4}
\usepackage{mathbbold}
\usepackage{amsfonts}
\usepackage{amsmath}
\usepackage{amssymb}
\usepackage{charter}
\usepackage{graphicx}

\begin{document}

\title{Two-mode light states before and after delocalized single-photon
addition}
\author{Bo Lan$^{1}$, Hong-chun Yuan$^{2}$ and Xue-xiang Xu$^{1,\dag }$}
\affiliation{$^{1}$College of Physics and Communication Electronics, Jiangxi Normal
University, Nanchang 330022, China;\\
$^{2}$School of Electrical and Information Engineering, Changzhou Institute
of Technology, Changzhou 213032, China\\
$^{\dag }$Corresponding author: xuxuexiang@jxnu.edu.cn }

\begin{abstract}
We studied the effect of delocalized single-photon addition (DPA $\hat{a}%
_{1}^{\dag }+e^{i\varphi }\hat{a}_{2}^{\dag }$) on two input modes
containing four cases: two independent coherent states (CSs), two
independent thermal states (TSs), two independent single-mode squeezed
vacuums (SVs), and an entangled two-mode squeezed vacuum (TMSV). In essence,
four types of non-Gaussian entangled light states are generated. We studied
three different resources (including entanglement, discorrelation and Wigner
negativity) for each two-mode light state. The output states after DPA are
entangled, with more parameters and complex structures, characterizing more
Wigner negativity or even discorrelation. In contrast, the CSs case is the
most tunable protocol, because its negativity under partial transposition,
discorrelation, and Wigner logarithmic negativity are more sensitive to
superposition phase $\varphi $ than those in TSs, SVs, and TMSV cases.


\end{abstract}

\maketitle

\section{Introduction}

Gaussian entangled states play important roles in continuous-variable (CV)
quantum physics\cite{1,2}. As the primary entangled resource, the two-mode
squeezed vacuum (TMSV) has been used in many quantum protocols, such as
quantum teleportation\cite{3} and quantum computation\cite{4}. However,
non-Gaussian states and operations are necessary for many quantum tasks\cite%
{5}. For example, the NOON state is a useful entangled resource, which has
been generated\cite{6} and applied in quantum metrology\cite{7}. The subject
of non-Gaussian quantum states has become a very active area of research. A
recent tutorial on non-Gaussian states was reported by Walschaers\cite{8}.
He provided a roadmap for the physics and an overview of several
experimental realizations of non-Gaussian quantum states. Lvovsky et al.
also reported another review, which covered theoretical and experimental
efforts to extend the applications from the Gaussian to the non-Gaussian
domain\cite{9}. In essence, non-Gaussian quantum states can be created by
applying non-Gaussian operations on the initial states. Moreover,
non-Gaussian operations, like single-photon subtraction (by annihilation
operator $\hat{a}$)\cite{10} and single-photon addition (by creation
operator $a^{\dag }$)\cite{11}, together with sequence and superpositions of
these two operations\cite{12},\ are essential to exploit quantum resources.
These resources (such as entanglement, discorrelation, and Wigner
negativity) can provide different quantum advantages in different quantum
protocols.

Many efforts have been devoted to classifying and quantifying entanglement
for bipartite systems\cite{13,14}. For example, one can measure pure-state
entanglement by entropy of entanglement\cite{15} and measure mixed-state
entanglement by entanglement of formation\cite{16}, entanglement cost\cite%
{17}, relative entropy of entanglement\cite{18}, and so on. Vidal and Werner
introduced a computable measure of entanglement for the bipartite mixed
state. They constructed negativity or a logarithmic negativity base on
negativity under partial transposition\textbf{\ }(NPT)\cite{19}. Later,
important developments were made in the optical hybrid approach to quantum
information\cite{20,21}. Morin et al. proposed and experimentally tested a
witness for single-photon entanglement\cite{22}. This witness specially
identified entanglement present in the single-photon subspace (\{$\left\vert
0\right\rangle $, $\left\vert 1\right\rangle $\}) and used it to witness
entanglement for CV quantum states\cite{23,24,25}.

Discorrelation, as a new joint statistical property of multimode quantum
light states, was introduced recently by Meyer-Scott et al.\cite{26}.
Discorrelation is characterized by the fact that the photon number in each
mode can take any values, but two modes together never exhibit the same.
Indeed, discorrelation can exhibit correlation of photon numbers between
different modes. Recently, Biagi et al. generated discorrelated states based
on delocalized photon addition (DPA)\cite{27,28} and provided experimental
observation and application\cite{29}. Indeed, discorrelation can be applied
to manipulating the secure distribution of information among untrusted
parties,\ in scenarios such as distributed voting schemes\cite{30} or
\textquotedblleft mental poker\textquotedblright\ games\cite{31}.

Wigner negativity is an essential resource to\ reaching a quantum advantage
with CVs\cite{32,33}. It has been identified as a necessary ingredient for
implementing processes that cannot be simulated efficiently with classical
resources\cite{4,34}. Wigner negativity is arguably one of the most striking
nonclassical and non-Gaussian features of quantum states. Wigner negativity
implies nonclassicality and non-Gaussianity. So far, researchers have
proposed many witnesses to quantify Wigner negativity\cite{35,36,37,38}.

In quantum-state engineering, there is a trend for researchers to prepare
multimode\cite{39} or multiphoton non-Gaussian quantum states\cite{40,41}.
In the study of states that are both highly non-Gaussian and highly
multimode, Walschaers et al. even provided a framework which is suited to
obtain general analytical results\cite{42}. Chabaud et al. also derived a
theoretical framework for the experimental certification of non-Gaussian
features of quantum states\cite{43}. Moreover, it has been demonstrated that
the entanglement of multimode states can be increased via local or nonlocal
operations\cite{44,45,46,47,48,49,50,51}. Indeed, nonlocal operations have
the effect of delocalization, which can entangle the input independent
states or change the entanglement of the input entangled states.

Then, what is the \textquotedblleft delocalized\textquotedblright\ quantum
operation? We consider that if $\hat{O}_{j}\equiv \hat{O}_{j}(\hat{a}%
_{j}^{\dag },\hat{a}_{j})$\ denotes the local quantum operation in the $j$%
-th mode, the coherent superpositions $\sum_{j=1}^{n}c_{j}\hat{O}_{j}$ are
referred to as a delocalized\ quantum operation [where $c_{j}$\ is the
superposition coefficient and $\hat{a}_{j}^{\dag }\ (\hat{a}_{j})$ is the
creation (annihilation) operator]. Undoubtedly, the effect of the coherent
superposition $\sum_{j=1}^{n}c_{j}\hat{O}_{j}$\ is different from that of
the local product $\hat{O}_{1}\hat{O}_{2}\cdots \hat{O}_{n}$. The
delocalized\ quantum operation can change properties of $n$-mode
(independent or entangled) light states and generate new $n$-mode entangled
light states. The possibility of arbitrarily \textquotedblleft
adding\textquotedblright\ and \textquotedblleft
subtracting\textquotedblright\ single photons to and from a light field may
give access to a complete engineering of quantum states\cite{52}. Photon
subtraction has been demonstrated as extremely useful for de-Gaussification%
\cite{53}, enhancing nonclassicality\cite{54}, and distilling entanglement%
\cite{55}. Ourjoumtsev et al. demonstrated that entanglement can be
increased via delocalized single-photon subtraction (DPS $c_{1}\hat{a}%
_{1}+c_{2}\hat{a}_{2}$)\cite{56}. Conversely, photon addition has been
demonstrated to create nonclassicality. Recently, Biagi et al. presented a
scheme to entangle two identical coherent states based on the delocalized
single-photon addition ($\hat{a}_{1}^{\dag }+e^{i\varphi }\hat{a}_{2}^{\dag }
$, where $\varphi $\ denotes superposition phase)\cite{27,28,29}. Indeed,
all these mode-selective photon additions and subtractions are
experimentally promising processes to create multimode non-Gaussian
entangled states\cite{57,58,59,60,61}.

In this paper, we theoretically study the effect of DPA (i.e, $\hat{a}%
_{1}^{\dag }+e^{i\varphi }\hat{a}_{2}^{\dag }$) on different input two-mode
states. We extend Biagi \textit{et al}.'s work\cite{27} to include
additional input states. Therefore, we generate entangled light states and
study their entanglement, discorrelation and Wigner negativity. The
remaining paper is organized as follows: In Sec.II, we introduce theoretical
schemes and entangled light states.\ In Sec.III, we witness entanglement for
two-mode light states. In Sec.IV, we study their discorrelation. In Sec.V,
we study their Wigner negativity. Main results are summarized and
discussions are given in Sec.VI.

\section{Entangled light states}

In 1991, Agarwal and Tara introduced the photon-added coherent state by
operating the creation operator\ on the coherent state\cite{62}. Since then,
a lot of similar works (including multimode cases) have been done one after
another\cite{63,64,65}. As depicted schematically in Fig.1, we consider four
types of non-Gaussian entangled light states by employing DPA $\hat{A}_{dl}=%
\hat{a}_{1}^{\dag }+e^{i\varphi }\hat{a}_{2}^{\dag }$ on different input
two-mode states, including two independent CSs, two independent thermal
states (TSs), two independent single-mode squeezed vacuums (SVs), and an
entangled TMSV.
\begin{figure}[tbp]
\label{Fig1} \centering\includegraphics[width=1.0\columnwidth]{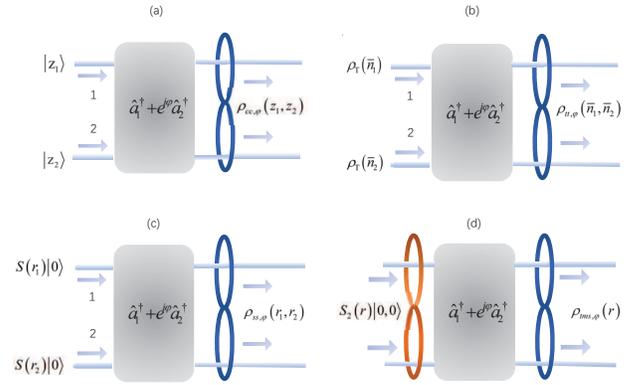}
\caption{{}Conceptual generating schemes of entangled light states $\protect%
\rho _{out,\protect\varphi }$ by performing DPA $a_{1}^{\dag }+e^{i\protect%
\varphi }a_{2}^{\dag }$ on different two-mode light states $\protect\rho %
_{in}$ including (a) two independent CSs, (b) two independent TSs, (c) two
independent SVs, and (d) an entangled TMSV.}
\end{figure}

\textit{Case CS-CS:} The CS in the $j$th mode is expressed as $\left\vert
z_{j}\right\rangle =e^{-\left\vert z_{j}\right\vert
^{2}/2}\sum_{n=0}^{\infty }\frac{z_{j}^{n}}{\sqrt{n!}}\left\vert
n\right\rangle $ (where $z_{j}$\ is an arbitrary complex number)\cite{66}.
Inputting two independent CSs (i.e., $\left\vert z_{1}\right\rangle $\ and $%
\left\vert z_{2}\right\rangle $), the generated state yields%
\begin{equation}
\left\vert \psi _{cc,\varphi }(z_{1},z_{2})\right\rangle _{12}=\frac{1}{%
\sqrt{N_{cc}}}\hat{A}_{dl}\left\vert z_{1}\right\rangle \left\vert
z_{2}\right\rangle   \label{II-1}
\end{equation}%
with normalization factor $N_{cc}=\left\vert z_{1}+e^{-i\varphi
}z_{2}\right\vert ^{2}+2$. In this case, the input state is $\rho
^{cc}(z_{1},z_{2})=\left\vert z_{1}\right\rangle \left\langle
z_{1}\right\vert \otimes \left\vert z_{2}\right\rangle \left\langle
z_{2}\right\vert $, and output state is $\rho _{cc,\varphi
}(z_{1},z_{2})=\left\vert \psi _{cc,\varphi }(z_{1},z_{2})\right\rangle
\left\langle \psi _{cc,\varphi }(z_{1},z_{2})\right\vert $. In particular,
when $z_{1}=z_{2}=\alpha $, the input is reduced to two identical CSs of
light in different modes, which is just the case studied in Ref.\cite%
{27,28,29}. When $z_{1}=0$ and $\alpha $, it will degenerate to the case in
other literature\cite{67,68}.

\textit{Case TS-TS:} The TS in the $j$th mode is diagonal in the Fock state
basis and is expressed as $\rho _{T}\left( \bar{n}_{j}\right)
=\sum_{n_{j}=0}^{\infty }P_{n_{j}}\left\vert n_{j}\right\rangle \left\langle
n_{j}\right\vert $ [where $P_{n_{j}}=\bar{n}_{j}{}^{n_{j}}/(\bar{n}%
_{j}+1)^{n_{j}+1}$ with mean thermal photon number $\bar{n}_{j}$]. Inputting
two independent TSs [i.e., the input state is $\rho ^{tt}(\bar{n}_{1},\bar{n}%
_{2})=\rho _{T}\left( \bar{n}_{1}\right) \otimes \rho _{T}\left( \bar{n}%
_{2}\right) $], the output state yields%
\begin{equation}
\rho _{tt,\varphi }(\bar{n}_{1},\bar{n}_{2})=\frac{1}{N_{tt}}\hat{A}%
_{dl}\rho _{T}\left( \bar{n}_{1}\right) \rho _{T}\left( \bar{n}_{2}\right)
\hat{A}_{dl}^{\dag }  \label{II-2}
\end{equation}%
with normalization factor $N_{tt}=\bar{n}_{1}+\bar{n}_{2}+2$. The TS is the
most classical light state and possesses a completely incoherent character.
Zavatta and coworkers have realized many light states by manipulating TSs%
\cite{69,70}.

\textit{Case SV-SV:} The SV in the $j$th mode is expressed as $\hat{S}\left(
r_{j}\right) \left\vert 0\right\rangle _{j}=(1-\lambda
_{j}^{2})^{1/4}e^{\lambda _{j}\hat{a}_{j}^{\dag 2}/2}\left\vert
0\right\rangle _{j}$ with $\lambda _{j}=\tanh r_{j}$ [where $S\left(
r_{j}\right) =e^{r_{j}(\hat{a}_{j}^{\dag 2}-\hat{a}_{j}^{2})/2}$ is the
single-mode squeezing operator with real $r_{j}$]\cite{71}. Inputting two
independent SVs [i.e., $\hat{S}\left( r_{1}\right) \left\vert 0\right\rangle
$ and $\hat{S}\left( r_{2}\right) \left\vert 0\right\rangle $], the
generated state yields%
\begin{equation}
\left\vert \psi _{ss,\varphi }(r_{1},r_{2})\right\rangle =\frac{1}{\sqrt{%
N_{ss}}}\hat{A}_{dl}\hat{S}\left( r_{1}\right) \left\vert 0\right\rangle
\hat{S}\left( r_{2}\right) \left\vert 0\right\rangle   \label{II-3}
\end{equation}%
with normalization factor $N_{ss}=\kappa _{1}+\kappa _{2}$ [here, $\kappa
_{j}=(1-\lambda _{j}^{2})^{-1}$]. In this case, the input state is $\rho
^{ss}(r_{1},r_{2})=\hat{S}\left( r_{1}\right) \left\vert 0\right\rangle
\left\langle 0\right\vert \hat{S}^{\dag }\left( r_{1}\right) \otimes \hat{S}%
\left( r_{2}\right) \left\vert 0\right\rangle \left\langle 0\right\vert \hat{%
S}^{\dag }\left( r_{2}\right) $ and the output state is $\rho _{ss,\varphi
}(r_{1},r_{2})=\left\vert \psi _{ss,\varphi }(r_{1},r_{2})\right\rangle
\left\langle \psi _{ss,\varphi }(r_{1},r_{2})\right\vert $.

\textit{Case TMSV:} The TMSV is written by $\hat{S}_{2}\left( r\right)
\left\vert 0_{1}0_{2}\right\rangle =\sqrt{1-\lambda ^{2}}e^{\lambda \hat{a}%
_{1}^{\dag }\hat{a}_{2}^{\dag }}\left\vert 0_{1}0_{2}\right\rangle $ with $%
\lambda =\tanh r$ [where $\hat{S}_{2}\left( r\right) =e^{r(\hat{a}_{1}^{\dag
}\hat{a}_{2}^{\dag }-\hat{a}_{1}\hat{a}_{2})}$ is the two-mode squeezing
operator with real $r$]\cite{72}. Inputting the TMSV, the generated state
yields%
\begin{equation}
\left\vert \psi _{tms,\varphi }(r)\right\rangle =\frac{1}{\sqrt{N_{tms}}}%
\hat{A}_{dl}\hat{S}_{2}\left( r\right) \left\vert 0_{1}0_{2}\right\rangle
\label{II-4}
\end{equation}%
with normalization factor $N_{tms}=2\kappa $ (here $\kappa =(1-\lambda
^{2})^{-1})$. In this case, the input state is $\rho ^{tms}(r)=\hat{S}%
_{2}\left( r\right) \left\vert 0_{1}0_{2}\right\rangle \left\langle
0_{1}0_{2}\right\vert \hat{S}_{2}^{\dag }\left( r\right) $, and the output
state is $\rho _{tms,\varphi }(r)=\left\vert \psi _{tms,\varphi
}(r)\right\rangle \left\langle \psi _{tms,\varphi }(r)\right\vert $. Unlike
the above three separable cases, the input is an entangled state in itself.

The input-output states in the above schemes can be unified as
\begin{equation}
\rho _{out,\varphi }=\frac{1}{N}\hat{A}_{dl}\rho _{in}\hat{A}_{dl}^{\dag }
\label{II-5}
\end{equation}%
with normalization factor $N$. Here $\rho _{in}$s include $\rho
^{cc}(z_{1},z_{2})$, $\rho ^{tt}(\bar{n}_{1},\bar{n}_{2})$, $\rho
^{ss}(r_{1},r_{2})$, and $\rho ^{tms}(r)$ and $\rho _{out,\varphi }$s
include $\rho _{cc,\varphi }(z_{1},z_{2})$, $\rho _{tt,\varphi }(\bar{n}_{1},%
\bar{n}_{2})$, $\rho _{ss,\varphi }(r_{1},r_{2})$ and $\rho _{tms,\varphi
}(r)$. Density operators are provided in Appendix A and normalization
factors are derived in Appendix B. In order to compare, we often simulate
properties by using the same $\bar{n}_{T,in}$ (input total mean photon
number), that is, $\left\vert z_{1}\right\vert ^{2}+\left\vert
z_{2}\right\vert ^{2}=\bar{n}_{T,in}$, $\bar{n}_{1}+\bar{n}_{2}=\bar{n}%
_{T,in}$, $\sinh ^{2}r_{1}+\sinh ^{2}r_{2}=\bar{n}_{T,in}$, and $2\sinh
^{2}r=\bar{n}_{T,in}$. When inputting $\left\vert 0\right\rangle
_{1}\left\vert 0\right\rangle _{2}$, the output will be a single-photon
mode-entangled state $\left\vert \psi _{00}\right\rangle =\frac{1}{\sqrt{2}}%
(\left\vert 1\right\rangle _{1}\left\vert 0\right\rangle _{2}$ $+e^{i\varphi
}\left\vert 0\right\rangle _{1}\left\vert 1\right\rangle _{2})$\cite{67}.
Once we let $z_{1}=z_{2}=0$, $\bar{n}_{1}=\bar{n}_{2}=0$, $r_{1}=r_{2}=0$ or
$r=0$, states in Eqs.(\ref{II-1})-(\ref{II-4}) are reduced to $\rho _{00}=$ $%
\left\vert \psi _{00}\right\rangle \left\langle \psi _{00}\right\vert $.

In what follows, we shall analyze entanglement, discorrelation and Wigner
negativity for all these two-mode light states.

\section{Entanglement of two-mode light states}

In this section, we shall quantify their entanglement by\ NPT\cite%
{13,14,19,73}, a witness proposed by Morin et al.\cite{22}. The remarkable
feature of this witness lie in dimension-independence for the measured
state. Here, we make a brief review of NPT.

First, we project measured state $\rho $ into subspace \{$\left\vert
00\right\rangle $, $\left\vert 01\right\rangle $, $\left\vert
10\right\rangle $ and $\left\vert 11\right\rangle $\} and obtain a $4\times 4
$ matrix with form%
\begin{equation}
X=\frac{1}{T}\left(
\begin{array}{cccc}
p_{00,00} & p_{00,01} & p_{00,10} & p_{00,11} \\
p_{01,00} & p_{01,01} & p_{01,10} & p_{01,11} \\
p_{10,00} & p_{10,01} & p_{10,10} & p_{10,11} \\
p_{11,00} & p_{11,01} & p_{11,10} & p_{11,11}%
\end{array}%
\right) ,  \label{III-1}
\end{equation}%
where $p_{k_{1}k_{2},l_{1}l_{2}}=\left\langle k_{1},k_{2}\right\vert \rho
\left\vert l_{1},l_{2}\right\rangle $ and $%
T=p_{00,00}+p_{01,01}+p_{10,10}+p_{11,11}$ (ensuring $\mathrm{Tr}X=1$)\cite%
{22,23,24,25}. Thus, the full density matrix was restricted to such
subspace. Here, $X$ is Hermitian and has non-negative eigenvalues, the trace
norm of which holds $\left\Vert X\right\Vert _{1}\equiv \mathrm{Tr}(\sqrt{%
X^{\dag }X})=\mathrm{Tr}X=1$\cite{19}.

Second, performing partial transposition in mode 2 for $X$, we obtain a
matrix with form
\begin{equation}
X^{T_{2}}=\frac{1}{T}\left(
\begin{array}{cccc}
p_{00,00} & p_{01,00} & p_{00,10} & p_{01,10} \\
p_{00,01} & p_{01,01} & p_{00,11} & p_{01,11} \\
p_{10,00} & p_{11,00} & p_{10,10} & p_{11,10} \\
p_{10,01} & p_{11,01} & p_{10,11} & p_{11,11}%
\end{array}%
\right) ,  \label{III-2}
\end{equation}%
with $X_{k_{1}l_{2},l_{1},k_{2}}^{T_{2}}=X_{k_{1}k_{2},l_{1},l_{2}}=\left%
\langle k_{1},k_{2}\right\vert X\left\vert l_{1},l_{2}\right\rangle $, and
further calculate eigenvalues of $X^{T_{2}}$. Certainly, $X^{T_{2}}$ also
satisfies $\mathrm{Tr}X^{T_{2}}=1$. In general, the trace norm of $X^{T_{2}}$
reads $\left\Vert X^{T_{2}}\right\Vert _{1}\equiv 1-2\sum_{i}\lambda _{i}^{-}
$, where $\lambda _{i}^{-}$ denotes negative eigenvalues of $X^{T_{2}}$\cite%
{19}.

Lastly, we quantify entanglement by NPT defined as%
\begin{equation}
\mathrm{NPT}(\rho )=-2\sum_{i}\lambda _{i}^{-},  \label{III-3}
\end{equation}%
where the factor of $-2$ ensures $0\leq \mathrm{NPT}(\rho )\leq 1$\cite%
{43,44,45}.

Indeed, NPT measures entanglement by considering how much $X^{T_{2}}$ fails
to be positive definite. If $X^{T_{2}}$\ has at least one negative
eigenvalue, $\rho $\ is inseparable (or entangled). NPT is zero for a
completely separable state (the $X^{T_{2}}$\ of which has no negative
eigenvalues) and $1$ for a maximally entangled one\cite{74}.

One can resort to $p_{k_{1}k_{2},l_{1},l_{2}}=\left\langle
k_{1},k_{2}\right\vert \rho \left\vert l_{1},l_{2}\right\rangle $ in
Appendix C and eigenvalues of $X^{T_{2}}$s in Appendix D to obtain NPTs.

(1) NPT of $\rho _{cc,\varphi }$ is calculated as
\begin{equation}
\mathrm{NPT}(\rho _{cc,\varphi }(z_{1},z_{2}))=\frac{2}{\left\vert
z_{1}+e^{-i\varphi }z_{2}\right\vert ^{2}+2}.  \label{III-4}
\end{equation}%
This result can be reduced to $\mathrm{NPT}(\rho _{cc,\varphi }(z,z))=$ $%
1/(1+\left\vert z\right\vert ^{2}(1+\cos \varphi ))$\cite{27}.

(2) NPT of $\rho _{tt,\varphi }$\ is calculated as%
\begin{equation}
\mathrm{NPT}(\rho _{tt,\varphi }(\bar{n}_{1},\bar{n}_{2}))=\frac{\sqrt{%
4A^{2}+\Gamma ^{2}}-\Gamma }{2A+\Gamma },  \label{III-5}
\end{equation}%
with $A=\left( 1+\overline{n}_{1}\right) \left( 1+\overline{n}_{2}\right) $
and $\Gamma =\overline{n}_{1}+\overline{n}_{2}+2\overline{n}_{1}\overline{n}%
_{2}$. Note that $\mathrm{NPT}(\rho _{tt,\varphi })$ is independent of $%
\varphi $.

(3) NPT of $\rho _{ss,\varphi }$ is calculated as%
\begin{equation}
\mathrm{NPT}(\rho _{ss,\varphi }(r_{1},r_{2}))=1.  \label{III-6}
\end{equation}%
This means that $\rho _{ss,\varphi }$ keeps maximum entanglement
independently of $r_{1}$, $r_{2}$, and $\varphi $.

(4) NPT of $\rho _{tms,\varphi }$ is calculated as%
\begin{equation}
\mathrm{NPT}(\rho _{tms,\varphi }(r))=1.  \label{III-7}
\end{equation}%
This means that $\rho _{tms,\varphi }$ keeps maximum entanglement
independently of $r$ and $\varphi $.
\begin{figure}[tbp]
\label{Fig2} \centering\includegraphics[width=1.0\columnwidth]{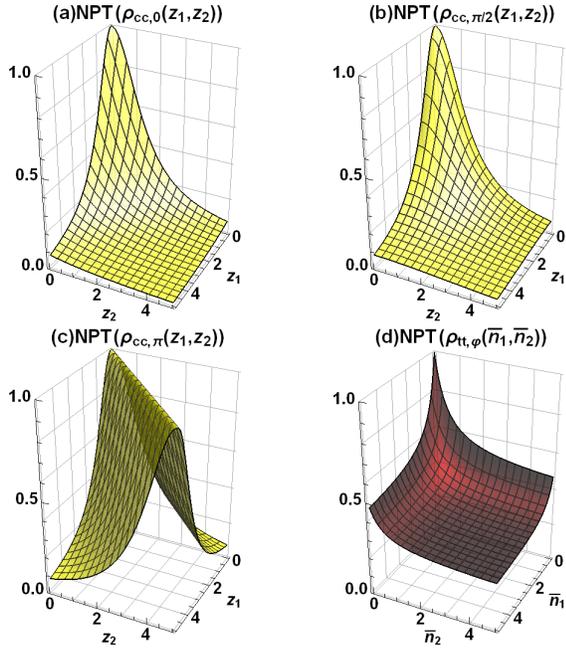}
\caption{Plots of $\mathrm{NPT}$s for (a) $\protect\rho _{cc,0}$, (b) $%
\protect\rho _{cc,\protect\pi /2}$, and (c) $\protect\rho _{cc,\protect\pi }$
in ($z_{1}$, $z_{2}$) space and (d) $\protect\rho _{tt,\protect\varphi }$ in
($\bar{n}_{1}$, $\bar{n}_{2}$) space.}
\end{figure}

In Fig.2, we plot $\mathrm{NPT}$s for $\rho _{cc,\varphi }$ in $(z_{1},z_{2})
$ space with different $\varphi $ $(0$, $\pi /2$, $\pi )$ and for $\rho
_{tt,\varphi }$ in $(\bar{n}_{1},\bar{n}_{2})$ space with arbitrary $\varphi
$. Without loss of generality, $z_{1}$\ and $z_{2}$ take real numbers. It is
noteworthy that $\mathrm{NPT}\left( \rho _{cc,\varphi }\right) $ can reach $1
$ at$\ z_{1}=z_{2}=0$ for any $\varphi $\ (see Fig.2(a),(b)) and $\mathrm{NPT%
}(\rho _{cc,\pi }(z,z))=1$ is always satisfied [see Fig.2(c)]. From
Fig.2(d), we see that $\mathrm{NPT}(\rho _{tt,\varphi })$ reaches$\mathrm{\ }%
1$ at$\ \bar{n}_{1}=\bar{n}_{2}=0$ and decreases as $\bar{n}_{1}$\ (or $\bar{%
n}_{2}$) increases. But $\mathrm{NPT}$s of $\rho _{ss,\varphi }$\ and $\rho
_{tms,\varphi }$\ are always $1$ independent of parameters. All cases can be
reduced to $\mathrm{NPT}(\rho _{00})=1$ as expected.

In Fig.3, we plot several NPTs as functions of $\bar{n}_{T,in}$.\ We find
that $\mathrm{NPT}(\rho _{cc,\pi }(z,z))$\textbf{, }$\mathrm{NPT}(\rho
_{ss,\varphi }(r_{1},r_{2}))$\ and $\mathrm{NPT}(\rho _{tms,\varphi }(r))$%
\textbf{\ }are always 1, showing their robust entanglements. These states
maintain constant maximum entanglement independently of parameters, while $%
\mathrm{NPT}(\rho _{cc,\varphi }(z,z))$s except $\varphi =\pi $ are monotone
decreasing functions of $\bar{n}_{T,in}$. For a given $\bar{n}_{T,in}$, the
larger $\varphi $ is, the larger $\mathrm{NPT}(\rho _{cc,\varphi }(z,z))$
is. Of course, $\mathrm{NPT}(\rho _{tt,\varphi }(\bar{n},\bar{n}))$ is also
a monotonically decreasing function of $\bar{n}_{T,in}$, which is
independent of $\varphi $.

\begin{figure}[tbp]
\label{Fig3} \centering\includegraphics[width=1.0\columnwidth]{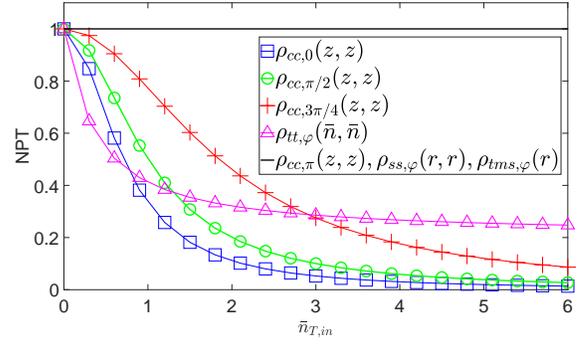}
\caption{$\mathrm{NPT}$s as\ functions of $\bar{n}_{T,in}$ for different $%
\protect\rho _{out,\protect\varphi }$s.}
\end{figure}

\section{Discorrelation of two-mode light states}

Discorrelation is used to show correlation of joint photon number
distributions (JPNDs). For two-mode light states, the JPND is defined by $%
P_{n_{1},n_{2}}=\left\langle n_{1},n_{2}\right\vert \rho \left\vert
n_{1},n_{2}\right\rangle $ (where $n_{j}$\ denotes photon number in mode $j$%
). If\ $P_{n,n}\equiv 0$ for all $n$, but $P_{n_{1},\cdot
}=\sum_{n_{2}=0}^{\infty }P_{n_{1},n_{2}}$ and $P_{\cdot
,n_{2}}=\sum_{n_{1}=0}^{\infty }P_{n_{1},n_{2}}$ are nonzero, then this
two-mode state is a discorrelated state\cite{26,27,28,29}. All JPNDs can be
calculated from $p_{k_{1}k_{2},l_{1}l_{2}}=\left\langle
k_{1},k_{2}\right\vert \rho \left\vert l_{1},l_{2}\right\rangle $ in
Appendix C.

JPNDs for $\rho _{cc,\pi }(z,z)$, $\rho _{tt,\varphi }(\bar{n},\bar{n})$, $%
\rho _{ss,\varphi }(r,r)$, and $\rho _{tms,\varphi }(r)$ with $\bar{n}%
_{T,in}=3$ are plotted in Fig.4. From Fig.4(a), we can affirm that only $%
\rho _{cc,\pi }(z,z)$\ is a discorrelated state because $P_{n,n}=0$ may hold
only if $z_{1}=z_{2}$ and $\varphi =\pi $. From Fig.4(b), we can affirm that
$\rho _{tt,\varphi }(\bar{n}_{1},\bar{n}_{2})$\ is not a discorrelated state
except $\bar{n}_{1}=\bar{n}_{2}=0$. As $\bar{n}_{1}$ and $\bar{n}_{2}$
increase, JPNDs of $\rho _{tt,\varphi }$ cannot meet characteristics of
discorrelation. From Fig.4(c), we can affirm that $\rho _{ss,\varphi
}(r_{1},r_{2})$ is a discorrelated state because $P_{n,n}\equiv 0$ is always
right regardless of $r_{1}$, $r_{2}$, and $\varphi $. From Fig.4(d), we can
affirm that $\rho _{tms,\varphi }(r)$ is a discorrelated state, which also
can be analyzed from twin-Fock distribution of TMSV\cite{75}. Of course, $%
\rho _{00}$\ is a special discorrelated state.
\begin{figure}[tbp]
\label{Fig4} \centering\includegraphics[width=1.0\columnwidth]{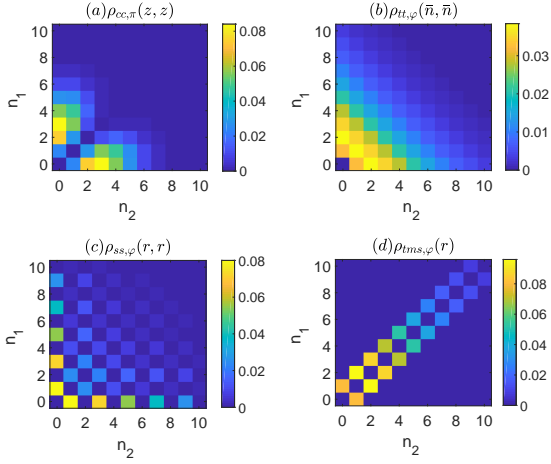}
\caption{JPNDs of states: (a) $\protect\rho _{cc,\protect\pi }(z,z)$, (b) $%
\protect\rho _{tt,\protect\varphi }(\bar{n},\bar{n})$, (c) $\protect\rho %
_{ss,\protect\varphi }(r,r)$, and (d) $\protect\rho _{tms,\protect\varphi %
}(r)$, with $\bar{n}_{T,in}=3$.}
\end{figure}

Figure 5 presents $P_{n,n}$s (with $n=0,1,2,3,4$) as functions of phase $%
\varphi $ for $\rho _{cc,\varphi }(\sqrt{1.5},\sqrt{1.5})$. As $\varphi $\
increases in interval $[0,\pi ]$, each $P_{n,n}$ (except $n=0$) decreases
monotonically. Until $\varphi =\pi $, we always have $P_{n,n}=0$, which is a
necessary condition of discorrelation. However, $\rho _{cc,\varphi }$s with $%
\varphi \neq \pi $ are not discorrelated states because $P_{n,n}=0$ is not
always satisfied for any nonzero $z_{1}$, $z_{2}$.
\begin{figure}[tbp]
\label{Fig5} \centering\includegraphics[width=0.9\columnwidth]{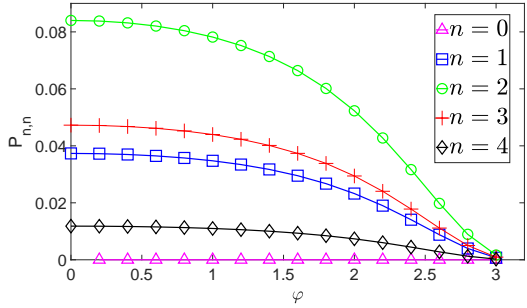}
\caption{$P_{n,n}$s as functions of phase $\protect\varphi $ for $\protect%
\rho _{cc,\protect\varphi }(\protect\sqrt{1.5},\protect\sqrt{1.5})$, with $%
P_{n,n}\equiv 0$ at $\protect\varphi =\protect\pi $.}
\end{figure}

\section{Wigner negativity of two-mode light states}

The Wigner function (WF) can exhibit negative values for some non-Gaussian
states. Wigner negativity, as a nonclassical indicator\cite{36,76,77,78}
with non-Gaussian character\cite{37,38,79}, plays an essential role in
quantum computing and simulation \cite{4,80}. Of course, Wigner negativity
can be detected \cite{81} or demonstrated \cite{82} in current technology.

The WF of a two-mode light state $\rho $ can be calculated as $W_{\rho
}\left( \beta _{1},\beta _{2}\right) =\left\langle \Pi \left( \beta
_{1}\right) \Pi \left( \beta _{2}\right) \right\rangle _{\rho }$. Here, $\Pi
\left( \beta _{j}\right) =\frac{2}{\pi }D(\beta _{j})(-1)^{\hat{a}_{j}^{\dag
}\hat{a}_{j}}D^{\dag }(\beta _{j})$ and $D(\beta _{j})=e^{\beta _{j}\hat{a}%
_{j}^{\dag }-\beta _{j}\hat{a}_{j}}$ denote the Wigner operator and
displacement operator for mode $j$, respectively, with $\beta _{j}=\left(
q_{j}+ip_{j}\right) /\sqrt{2}$. Obviously, $W_{\rho }\left( \beta _{1},\beta
_{2}\right) $ is real and bounded by $-4/\pi ^{2}\leq W_{\rho }\left( \beta
_{1},\beta _{2}\right) \leq 4/\pi ^{2}$.

Interestingly, $W_{\rho _{in}}$\ and $W_{\rho _{out,\varphi }}$ can be
linked by%
\begin{equation}
W_{\rho _{out,\varphi }}\left( \beta _{1},\beta _{2}\right) =\frac{T^{NG}}{N}%
W_{\rho _{in}}\left( \beta _{1},\beta _{2}\right) ,  \label{X-1}
\end{equation}%
where $T^{NG}$ and $N$\ denote the non-Gaussian term and normalization
factor, respectively. All analytical expressions can be obtained from
Appendix E. The results show that all $W_{\rho _{in}}$s are Gaussian and all
$W_{\rho _{out,\varphi }}$s are non-Gaussian. If condition $T^{NG}<0$ is
satisfied, the negativity of $W_{\rho _{out,\varphi }}$\ is exhibited.

\textit{Case CS-CS:} We can obtain $W_{\rho _{cc,\varphi }}$ after knowing
\begin{equation}
W_{\rho ^{cc}}=\frac{4e^{-2\left\vert z_{1}-\beta _{1}\right\vert
^{2}-2\left\vert z_{2}-\beta _{2}\right\vert ^{2}}}{\pi ^{2}},  \label{X-2}
\end{equation}%
and%
\begin{equation}
T_{cc}^{NG}=|(z_{1}-2\beta _{1})+e^{-i\varphi }(z_{2}-2\beta _{2})|^{2}-2.
\label{X-3}
\end{equation}

\textit{Case TS-TS:} We can obtain $W_{\rho _{tt,\varphi }}$ after knowing
\begin{equation}
W_{\rho ^{tt}}=\frac{4e^{-\allowbreak \frac{2}{2\bar{n}_{1}+1}\left\vert
\beta _{1}\right\vert ^{2}-\allowbreak \frac{2}{2\bar{n}_{2}+1}\left\vert
\beta _{2}\right\vert ^{2}}}{\pi ^{2}\left( 2\bar{n}_{1}+1\right) \left( 2%
\bar{n}_{2}+1\right) }  \label{X-4}
\end{equation}%
and%
\begin{equation}
T_{tt}^{NG}=4\left\vert \epsilon _{1}\beta _{1}+e^{-i\varphi }\epsilon
_{2}\beta _{2}\right\vert ^{2}-(\epsilon _{1}+\epsilon _{2})  \label{X-5}
\end{equation}%
\ with $\epsilon _{j}=(\bar{n}_{j}+1)/(2\bar{n}_{j}+1)$.

\textit{Case SV-SV:} We can obtain $W_{\rho _{ss,\varphi }}$ after knowing
\begin{equation}
W_{\rho ^{ss}}=\frac{4e^{-2\kappa _{1}\left\vert {}\allowbreak \lambda
_{1}\beta _{1}-\beta _{1}^{\ast }\right\vert ^{2}-2\kappa _{2}\left\vert
{}\allowbreak \lambda _{2}\beta _{2}-\beta _{2}^{\ast }\right\vert ^{2}}}{%
\pi ^{2}}  \label{X-6}
\end{equation}%
and%
\begin{eqnarray}
T_{ss}^{NG} &=&4|\kappa _{1}(\lambda _{1}\beta _{1}-{}\allowbreak \beta
_{1}^{\ast })+e^{-i\varphi }\kappa _{2}({}\allowbreak \lambda _{2}\beta
_{2}-\beta _{2}^{\ast })|^{2}  \notag \\
&&-(\kappa _{1}+\kappa _{2}).  \label{X-7}
\end{eqnarray}

\textit{Case TMSV:} We can obtain $W_{\rho _{tms,\varphi }}$ after knowing
\begin{equation}
W_{\rho ^{tms}}=\frac{4}{\pi ^{2}}e^{-2\kappa (\left\vert \lambda \beta
_{1}-\beta _{2}^{\ast }\right\vert ^{2}+\left\vert \lambda {}\allowbreak
\beta _{2}-\beta _{1}^{\ast }\right\vert ^{2})}  \label{X-8}
\end{equation}%
and%
\begin{eqnarray}
T_{tms}^{NG} &=&4\kappa ^{2}|(\lambda \beta _{1}-\beta _{2}^{\ast
})+e^{-i\varphi }({}\allowbreak \lambda \beta _{2}-\beta _{1}^{\ast })|^{2}
\notag \\
&&-2\kappa .  \label{X-9}
\end{eqnarray}

Using the above expressions, we can obtain $W_{\rho _{out,\varphi }}$ for
each case, which can be reduced to the following extreme case
\begin{equation}
W_{\rho _{00}}=\frac{4(2\left\vert \beta _{1}+e^{-i\varphi }\beta
_{2}\right\vert ^{2}-1)}{\pi ^{2}}e^{-\allowbreak 2(\left\vert \beta
_{1}\right\vert ^{2}+\left\vert \beta _{2}\right\vert ^{2})}.  \label{X-10}
\end{equation}

The distribution $W_{\rho }\left( \beta _{1},\beta _{2}\right) $ can be
expressed in the four-dimensional phase space as $W_{\rho }\left(
q_{1},p_{1};q_{2},p_{2}\right) $, which shows correlations existing in four
quadratures ($\hat{q}_{1}$, $\hat{p}_{1}$, $\hat{q}_{2}$, $\hat{p}_{2}$).
Here, $\hat{q}_{j}=(\hat{a}_{j}+\hat{a}_{j}^{\dag })/\sqrt{2}$ and $\hat{p}%
_{j}=(\hat{a}_{j}-\hat{a}_{j}^{\dag })/(i\sqrt{2})$ denote the position
operator and momentum operator of mode $j$, respectively. $W_{\rho
_{00}}\left( \beta _{1},\beta _{2}\right) $ was pictorially demonstrated by
Ourjoumtsev \textit{et al}.\cite{56}. Figure.6 presents contour plots of $%
W_{\rho }\left( q_{1},0;q_{2},0\right) $ in the ($q_{1}$, $q_{2}$) section
for each case with one-input $\rho _{in}$ and three-output $\rho
_{out,\varphi }$ ($\varphi =0,\pi /2,\pi $). Before DPA, the input WFs are
Gaussian without Wigner negativity. But after DPA, the output WFs will be
non-Gaussian with Wigner negativity.
\begin{figure*}[tbp]
\label{Fig6} \centering\includegraphics[width=2.0\columnwidth]{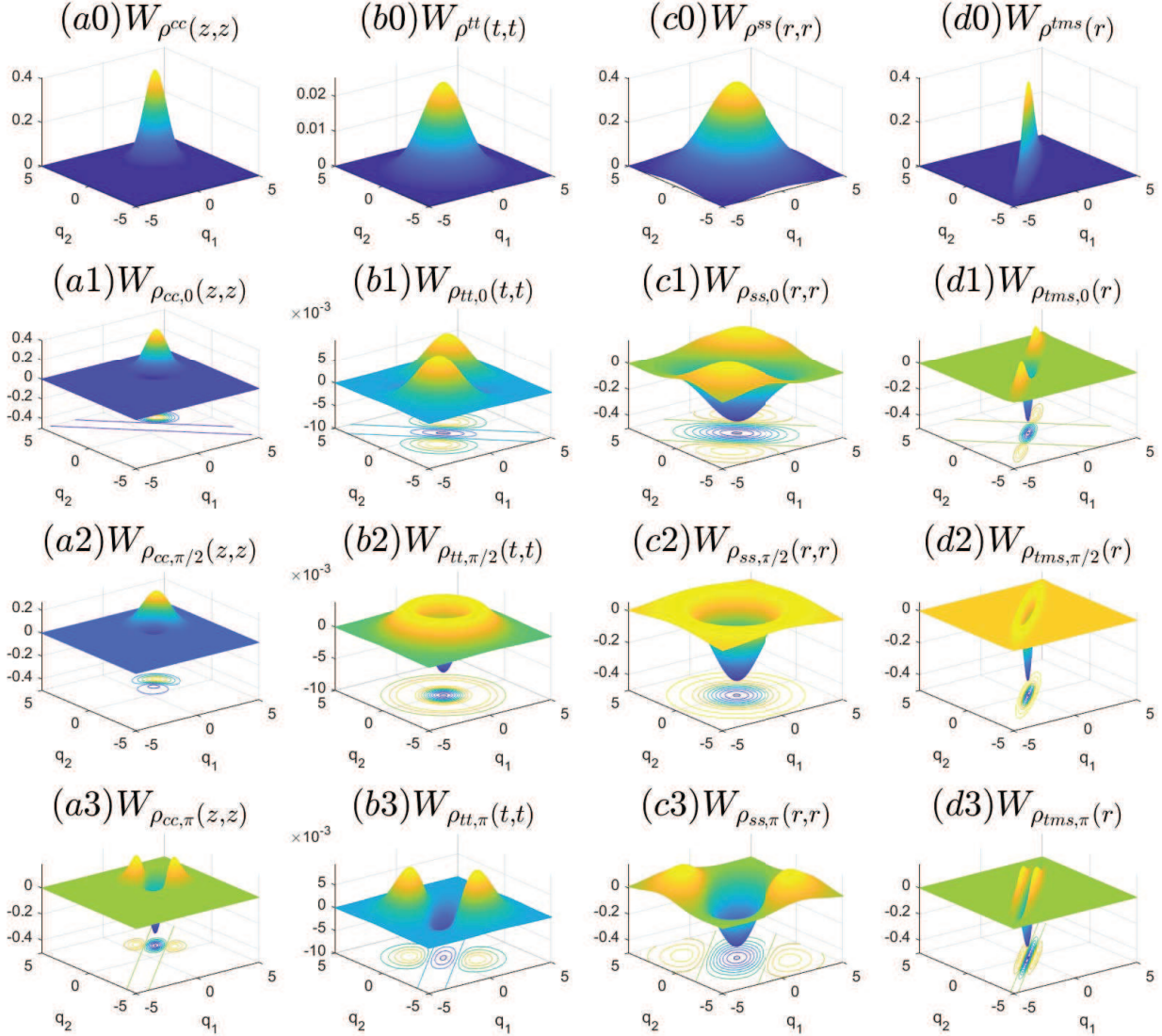}
\caption{Plots of $W_{\protect\rho }\left( q_{1},0;q_{2},0\right) $ in $%
\left( q_{1},q_{2}\right) $ space for (a) $\protect\rho ^{cc}$ and $\protect%
\rho _{cc,\protect\varphi }$ with $\protect\varphi =0$, $\protect\pi /2$, $%
\protect\pi $; (b) $\protect\rho ^{tt}$ and $\protect\rho _{tt,\protect%
\varphi }$ with $\protect\varphi =0$, $\protect\pi /2$, $\protect\pi $; (c) $%
\protect\rho ^{ss}$ and $\protect\rho _{ss,\protect\varphi }$ with $\protect%
\varphi =0$, $\protect\pi /2$, $\protect\pi $; and (d) $\protect\rho ^{tms}$
and $\protect\rho _{tms,\protect\varphi }$ with $\protect\varphi =0$, $%
\protect\pi /2$, $\protect\pi $. Here, their inputs are symmetrical with $%
\bar{n}_{T,in}=3$. }
\end{figure*}

From a quantitative perspective, we can measure Wigner negativity by
quantifying Wigner logarithmic negativity (WLN)\cite{37,38}. Here, the WLN
is defined by%
\begin{equation}
\mathrm{WLN}=\ln (\int \left\vert W_{\rho }\left( \beta _{1},\beta
_{2}\right) \right\vert d^{2}\beta _{1}d^{2}\beta _{2}),  \label{X-A}
\end{equation}%
which shows non-Gaussianity of the quantum state. By fixing $\bar{n}_{T,in}=1
$, $2$, and $3$, we plot WLN behaviors as functions of $\varphi $ in the
interval $[0,\pi ]$ for different $\rho _{out,\varphi }$s in Fig.7. The
results show the following:

(1) WLNs of $\rho _{cc,\varphi }$s\ are monotone increasing functions of $%
\varphi $. For symmetrical $z_{1}=z_{2}$ cases, the maximum WLNs are equal
to $0.35$ at $\varphi =\pi $, while for asymmetrical $z_{1}\neq z_{2}$
cases, the maximum WLNs are smaller than $0.35$ at $\varphi =\pi $.

(2) WLNs of $\rho _{tt,\varphi }$s are independent of $\varphi $ but
determined by $\bar{n}_{1}\ $and $\bar{n}_{2}$. The larger $\bar{n}_{T,in}$\
is, the smaller WLN\ is. As $\bar{n}_{T,in}$\ increases, the WLNs limit to
zero. Moreover, WLN for the asymmetrical case is bigger than that for the
symmetrical case for fixing $\bar{n}_{T,in}$.

(3) WLNs of $\rho _{ss,\varphi }(r)$ will remain at $0.35$ for any $r_{1}$, $%
r_{2}$, and $\varphi $.

(4) WLNs of $\rho _{ss,\varphi }(r)$ will be about $0.35$ for any $r\ $and $%
\varphi $. Indeed WLN of $\rho _{00}$\ is always $0.35$ for any $\varphi $.
\begin{figure}[tbp]
\label{Fig7} \centering\includegraphics[width=1.0\columnwidth]{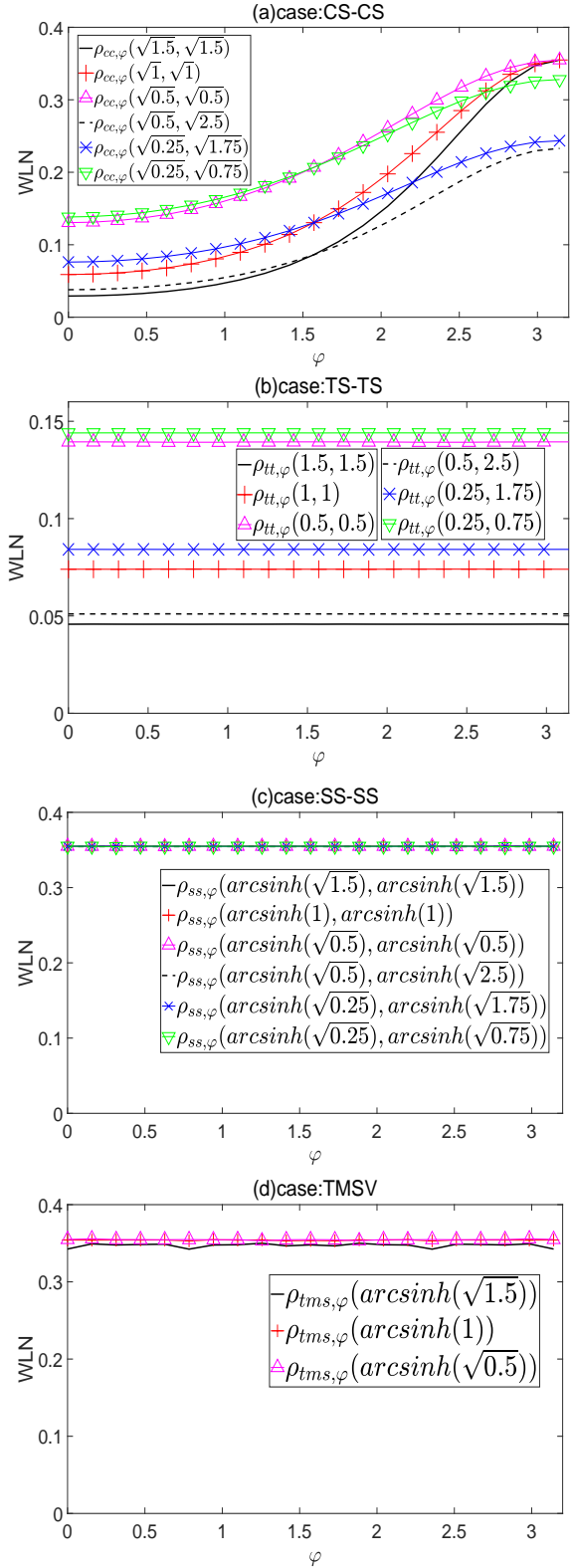}
\caption{ WLNs as functions of $\protect\varphi $ for states (a) $\protect%
\rho _{cc,\protect\varphi }(z_{1},z_{2})$ with different $z_{1}$\ and $z_{2}$%
, where the maximal WLN values at $\protect\varphi =\protect\pi $ are
corresponding to 0.35, 0.328, 0.244, and 0.233 from top to bottom; (b) $%
\protect\rho _{tt,\protect\varphi }(\bar{n}_{1},\bar{n}_{2})$ with different
$(\bar{n}_{1},\bar{n}_{2})$, where the WLN values are corresponding to
0.144, 0.139, 0.084, 0.074, 0.051, and 0.046 from top to bottom; (c) $%
\protect\rho _{ss,\protect\varphi }(r_{1},r_{2})$ with different $r_{1}$\
and $r_{2}$, where the WLN values are equal to 0.35; and (d) $\protect\rho %
_{tms,\protect\varphi }(r)$ with different $r$, where the WLN values are
equal to or slightly smaller than 0.35.}
\end{figure}

\section{Conclusion and discussion}

In summary, we have studied NPT, discorrelation, and WLN for four different
two-mode states subjected to a DPA operator. The four two-mode states are
two independent CSs, two independent TSs, two independent SVs, and a TMSV,
which are the common quantum states in the field of quantum optics. We
calculated that all four of these states lead to non-Gaussian entangled
states following DPA. Based on NPTs, we concluded entanglement in Table I.
Based on proper criteria, we concluded discorrelation in Table II. We also
calculated WLNs to exhibit Wigner negativity. In contrast, we still think
that the CSs protocol is the most tunable. This is because, NPT,
discorrelation, and WLN in the CSs case are all very sensitive to $\varphi $%
\ and reach maximal values (or effects) at $\varphi =\pi $. So $\rho
_{cc,\varphi }$s have good controllability by adjusting $\varphi $,\ while\
in TSs, SVs, and TMSV protocols, NPT, discorrelation, and WLN are
independent of $\varphi $. Moreover, SVs and TMSV after DPA do not exhibit
more entanglement and Wigner negativity than $\left\vert 0\right\rangle
_{1}\left\vert 0\right\rangle _{2}$\ after DPA, which can be seen from their
NPTs and WLNs. Worst of all, as thermal photon number increases, TSs after
DPA have lower NPT and WLN than $\left\vert 0\right\rangle _{1}\left\vert
0\right\rangle _{2}$\ after DPA, with discorrelation disappearance.
\begin{table}[h]
\caption{Results of NPT and entanglement for all two-mode light states
considered in this paper.}
\begin{center}
\begin{tabular}{|c||c|c|}
\hline\hline
& NPT & Entangled or Separable \\ \hline\hline
$\rho ^{cc}(z_{1},z_{2})$ & $0$ & Separable \\ \hline
$\rho _{cc,\varphi }(z_{1},z_{2})$ & $\frac{2}{\left\vert z_{1}+e^{-i\varphi
}z_{2}\right\vert ^{2}+2}$ & Entangled \\ \hline
$\rho ^{tt}(\bar{n}_{1},\bar{n}_{2})$ & $0$ & Separable \\ \hline
$\rho _{tt,\varphi }(\bar{n}_{1},\bar{n}_{2})$ & $\frac{\sqrt{4A^{2}+\Gamma
^{2}}-\Gamma }{2A+\Gamma }$ & Entangled \\ \hline
$\rho ^{ss}(r_{1},r_{2})$ & $0$ & Separable \\ \hline
$\rho _{ss,\varphi }(r_{1},r_{2})$ & $1$ & Entangled \\ \hline
$\rho ^{tms}(r)$ & $\frac{2\lambda }{1+\lambda ^{2}}$ & Entangled \\ \hline
$\rho _{tms,\varphi }(r)$ & $1$ & Entangled \\ \hline
\end{tabular}%
\end{center}
\end{table}
\begin{table}[h]
\caption{Results of discorrelation for all two-mode light states considered
in this paper.}
\begin{center}
\begin{tabular}{|c||c|c|}
\hline\hline
& Discorrelation & Conditions \\ \hline\hline
$\rho ^{cc}(z_{1},z_{2})$ & No & for arbitrary $z_{1},$ $z_{2}$ \\ \hline
$\rho _{cc,\varphi }(z_{1},z_{2})$ & Yes & only if $z_{1}=z_{2}$, and $%
\varphi =\pi $ \\ \hline
$\rho ^{tt}(\bar{n}_{1},\bar{n}_{2})$ & No & for arbitrary $\bar{n}_{1},\bar{%
n}_{2}$ \\ \hline
$\rho _{tt,\varphi }(\bar{n}_{1},\bar{n}_{2})$ & No & for nonzero $\bar{n}%
_{1},\bar{n}_{2}$ and $\varphi $ \\ \hline
$\rho ^{ss}(r_{1},r_{2})$ & No & for arbitrary $r_{1},r_{2}$ \\ \hline
$\rho _{ss,\varphi }(r_{1},r_{2})$ & Yes & for arbitrary $r_{1},r_{2}$ and $%
\varphi $ \\ \hline
$\rho ^{tms}(r)$ & No & for arbitrary $r$ \\ \hline
$\rho _{tms,\varphi }(r)$ & Yes & for arbitrary $r$ and $\varphi $ \\ \hline
\end{tabular}%
\end{center}
\end{table}

We regret that our schemes are ideal without considering decoherence
(including losses or detection efficiency). Recently, Walschaers et al. used
the framework of open quantum systems and master equations to describe
losses in all modes of the considered protocol\cite{83}. This provided us a
reference for our future works. Rather than changing input states, we can
also generalize $\hat{a}_{1}^{\dag }+e^{i\varphi }\hat{a}_{2}^{\dag }$ to
multiphoton cases, such as $\hat{a}_{1}^{\dag m_{1}}+e^{i\varphi }\hat{a}%
_{2}^{\dag m_{2}}$ or $(\hat{a}_{1}^{\dag }+e^{i\varphi }\hat{a}_{2}^{\dag
})^{m}$. Broadly, $\hat{a}_{1}^{\dag }+e^{i\varphi }\hat{a}_{2}^{\dag }$ can
be extended to $c_{1}\hat{O}_{1}(\hat{a}_{1}^{\dag },\hat{a}_{1})+c_{2}\hat{O%
}_{2}(\hat{a}_{2}^{\dag },\hat{a}_{2})$, including $\hat{a}_{1}+e^{i\varphi }%
\hat{a}_{2}$. In addition, we can extend from two-mode to multimode cases.

So far, in relation to applications or quantum technologies, many
breakthroughs of delocalized photon addition or subtraction were reported.
Roeland et al. developed a general framework for mode-selective
single-photon addition to a multimode quantum field\cite{84}. Barral and
Linares proposed a versatile integrated optical chip and generated both
non-local addition and subtraction on SVs\cite{85}. Barral et al. produced a
reconfigurable superposition of photon subtraction on two SVs\cite{86}.
Nehra et al. produced a variety of non-Gaussian states by using coherent
photon subtraction from a TMSV\cite{87}. So, we also anticipate that our
considered states will be generated in state-of-the-art quantum optics
experiments and will be used as resources in quantum technologies.

\section*{Appendix: Supplemental materials}

Here we provide supplemental materials for the main text.

\subsection*{Appendix A: Unifying expressions of density operators}

Obviously, once we assume
\begin{subequations}
\begin{equation}
\rho _{h_{1},h_{2},t_{1},t_{2}}=\hat{a}_{1}^{\dag h_{1}}\hat{a}_{2}^{\dag
h_{2}}\rho _{in}\hat{a}_{1}^{t_{1}}\hat{a}_{2}^{t_{2}},  \tag{A1}
\end{equation}%
(here, $h_{1}$, $h_{2}$, $t_{1}$, and $t_{2}$ are non-negative integers), we
can obtain $\rho _{in}$ by using $\rho _{in}=\rho _{0,0,0,0}$ and $\rho
_{out,\varphi }$\ by using
\end{subequations}
\begin{subequations}
\begin{equation}
\rho _{out,\varphi }=N^{-1}(\rho _{1,0,1,0}+\rho _{0,1,0,1}+e^{-i\varphi
}\rho _{1,0,0,1}+e^{i\varphi }\rho _{0,1,1,0}).  \tag{A2}
\end{equation}%
In the following calculation, we rewrite $\rho _{h_{1},h_{2},t_{1},t_{2}}$
as
\end{subequations}
\begin{subequations}
\begin{align}
\rho _{h_{1},h_{2},t_{1},t_{2}}& =\partial _{\mu _{1}}^{h_{1}}\partial _{\mu
_{2}}^{h_{2}}\partial _{\nu _{1}}^{t_{1}}\partial _{\nu _{2}}^{t_{2}}e^{\mu
_{1}\hat{a}_{1}^{\dag }}e^{\mu _{2}\hat{a}_{2}^{\dag }}\rho _{in}  \notag \\
& e^{\nu _{1}\hat{a}_{1}}e^{\nu _{2}\hat{a}_{2}}|_{\mu _{1}=\mu
_{2}=v_{1}=\nu _{2}=0},  \tag{A3}
\end{align}%
where $a_{j}^{\dag h_{j}}=\partial _{\mu _{j}}^{h_{j}}e^{\mu _{j}\hat{a}%
_{j}^{\dag }}|_{\mu _{j}=0}$ and $a_{j}^{t_{j}}=\partial _{\nu
_{j}}^{t_{j}}e^{\nu _{j}\hat{a}_{j}}|_{v_{j}=0}$ have been used.

\subsection*{Appendix B: Normalization factors of $\protect\rho _{out,%
\protect\varphi }$}

As long as we know $\mathrm{Tr}\rho _{h_{1},h_{2},t_{1},t_{2}}$, we can
obtain normalization factor $N$ for $\rho _{out,\varphi }$ by
\end{subequations}
\begin{subequations}
\begin{equation}
N=\mathrm{Tr}\rho _{1,0,1,0}+\mathrm{Tr}\rho _{0,1,0,1}+e^{-i\varphi }%
\mathrm{Tr}\rho _{1,0,0,1}+e^{i\varphi }\mathrm{Tr}\rho _{0,1,1,0}.  \tag{B1}
\end{equation}%
Next we give $\mathrm{Tr}\rho _{h_{1},h_{2},t_{1},t_{2}}$\ for each case.

(I) For case CS-CS, we have
\end{subequations}
\begin{subequations}
\begin{align}
\mathrm{Tr}\rho _{h_{1},h_{2},t_{1},t_{2}}^{cc}& =\partial _{\mu
_{1}}^{h_{1}}\partial _{\mu _{2}}^{h_{2}}\partial _{\nu
_{1}}^{t_{1}}\partial _{\nu _{2}}^{t_{2}}e^{\mu _{1}\nu _{1}+\nu
_{1}z_{1}+\mu _{1}z_{1}^{\ast }}  \notag \\
& e^{\mu _{2}\nu _{2}+\nu _{2}z_{2}+\mu _{2}z_{2}^{\ast }\allowbreak }|_{\mu
_{1}=\mu _{2}=v_{1}=\nu _{2}=0},  \tag{B2}
\end{align}%
which determines $\mathrm{Tr}\rho _{1,0,1,0}^{cc}=\left\vert \allowbreak
z_{1}\right\vert ^{2}+1$, $\mathrm{Tr}\rho _{0,1,0,1}^{cc}=\left\vert
\allowbreak z_{2}\right\vert ^{2}+1$, $\mathrm{Tr}\rho
_{1,0,0,1}^{cc}=z_{1}^{\ast }\allowbreak z_{2}$, $\mathrm{Tr}\rho
_{0,1,1,0}^{cc}=\allowbreak z_{1}z_{2}^{\ast }$, and $N_{cc}=\left\vert
z_{1}+e^{-i\varphi }z_{2}\right\vert ^{2}+2$.

(II) For case TS-TS, we have
\end{subequations}
\begin{subequations}
\begin{align}
\mathrm{Tr}\rho _{h_{1},h_{2},t_{1},t_{2}}^{tt}& =\partial _{\mu
_{1}}^{h_{1}}\partial _{\mu _{2}}^{h_{2}}\partial _{\nu
_{1}}^{t_{1}}\partial _{\nu _{2}}^{t_{2}}e^{\left( \bar{n}_{1}+1\right) \mu
_{1}\nu _{1}}  \notag \\
& e^{\left( \bar{n}_{2}+1\right) \mu _{2}\nu _{2}}|_{\mu _{1}=\mu
_{2}=v_{1}=\nu _{2}=0},  \tag{B3}
\end{align}%
which determines $\mathrm{Tr}\rho _{1,0,1,0}^{tt}=\bar{n}_{1}+1$, $\mathrm{Tr%
}\rho _{0,1,0,1}^{tt}=\bar{n}_{2}+1$, $\mathrm{Tr}\rho _{1,0,0,1}^{tt}=0$, $%
\mathrm{Tr}\rho _{0,1,1,0}^{tt}=0$, and $N_{tt}=\bar{n}_{1}+\bar{n}_{2}+2$.

(III) For case SV-SV, we have
\end{subequations}
\begin{subequations}
\begin{align}
\mathrm{Tr}\rho _{h_{1},h_{2},t_{1},t_{2}}^{ss}& =\partial _{\mu
_{1}}^{h_{1}}\partial _{\mu _{2}}^{h_{2}}\partial _{\nu
_{1}}^{t_{1}}\partial _{\nu _{2}}^{t_{2}}e^{\kappa _{1}\mu _{1}\nu
_{1}+\lambda _{1}\kappa _{1}(\nu _{1}^{2}+\mu _{1}^{2})/2}  \notag \\
& e^{\kappa _{2}\mu _{2}\nu _{2}+\lambda _{2}\kappa _{2}(\nu _{2}^{2}+\mu
_{2}^{2})/2}|_{\mu _{1}=\mu _{2}=v_{1}=\nu _{2}=0},  \tag{B4}
\end{align}%
which determines $\mathrm{Tr}\rho _{1,0,1,0}^{ss}=\kappa _{1}$, $\mathrm{Tr}%
\rho _{0,1,0,1}^{ss}=\kappa _{2}$, $\mathrm{Tr}\rho _{1,0,0,1}^{ss}=0$, $%
\mathrm{Tr}\rho _{0,1,1,0}^{ss}=0$, and $N_{ss}=\kappa _{1}+\kappa _{2}$.

(IV) For case TMSV, we have
\end{subequations}
\begin{subequations}
\begin{align}
\mathrm{Tr}\rho _{h_{1},h_{2},t_{1},t_{2}}^{tms}& =\partial _{\mu
_{1}}^{h_{1}}\partial _{\mu _{2}}^{h_{2}}\partial _{\nu
_{1}}^{t_{1}}\partial _{\nu _{2}}^{t_{2}}e^{\kappa \left( \mu _{1}\nu
_{1}+\allowbreak \mu _{2}\nu _{2}\right) }  \notag \\
& e^{\lambda \kappa \left( \mu _{1}\mu _{2}+\nu _{1}\nu _{2}\right) }|_{\mu
_{1}=\mu _{2}=v_{1}=\nu _{2}=0},  \tag{B5}
\end{align}%
which determines $\mathrm{Tr}\rho _{1,0,1,0}^{tms}=\mathrm{Tr}\rho
_{0,1,0,1}^{tms}=\kappa $, $\mathrm{Tr}\rho _{1,0,0,1}^{12}=\mathrm{Tr}\rho
_{0,1,1,0}^{12}=0$, and $N_{tms}=2\kappa $.

\subsection*{Appendix C: Density-matrix elements in Fock basis}

If we know

\end{subequations}
\begin{equation}
p_{h_{1},h_{2},t_{1},t_{2},k_{1}k_{2},l_{1}l_{2}}=\left\langle
k_{1}\right\vert \left\langle k_{2}\right\vert \rho
_{h_{1},h_{2},t_{1},t_{2}}\left\vert l_{1}\right\rangle \left\vert
l_{2}\right\rangle ,  \tag{C1}
\end{equation}%
we can easily obtain $%
p_{k_{1}k_{2},l_{1}l_{2}}^{in}=p_{0,0,0,0,k_{1}k_{2},l_{1}l_{2}}$ for $\rho
_{in}$ and
\begin{subequations}
\begin{align}
& p_{k_{1}k_{2},l_{1}l_{2}}^{out}  \notag \\
& =N^{-1}(p_{1,0,1,0,k_{1}k_{2},l_{1}l_{2}}+p_{0,1,0,1,k_{1}k_{2},l_{1}l_{2}}
\notag \\
& +e^{-i\varphi }p_{1,0,0,1,k_{1}k_{2},l_{1}l_{2}}+e^{i\varphi
}p_{0,1,1,0,k_{1}k_{2},l_{1}l_{2}})  \tag{C2}
\end{align}%
for $\rho _{out,\varphi }$. By employing $\left\langle k_{j}\right\vert =%
\frac{1}{\sqrt{k_{j}!}}\partial _{m_{j}}^{k_{j}}\left\langle 0\right\vert
e^{m_{j}\hat{a}_{j}}|_{m_{j}=0}$ and $\left\vert l_{j}\right\rangle =\frac{1%
}{\sqrt{l_{j}!}}\partial _{n_{j}}^{l_{j}}e^{n_{j}\hat{a}_{j}^{\dag
}}\left\vert 0\right\rangle |_{n_{j}=0}$, we have
\end{subequations}
\begin{subequations}
\begin{align}
& p_{h_{1},h_{2},t_{1},t_{2},k_{1}k_{2},l_{1}l_{2}}  \notag \\
& =\frac{1}{\sqrt{k_{1}!k_{2}!l_{1}!l_{2}!}}\partial
_{m_{1}}^{k_{1}}\partial _{m_{2}}^{k_{2}}\partial _{n_{1}}^{l_{1}}\partial
_{n_{2}}^{l_{2}}\partial _{\mu _{1}}^{h_{1}}\partial _{\mu
_{2}}^{h_{2}}\partial _{\nu _{1}}^{t_{1}}\partial _{\nu _{2}}^{t_{2}}  \notag
\\
& \left\langle 0\right\vert _{1}\left\langle 0\right\vert _{2}e^{m_{1}\hat{a}%
_{1}}e^{m_{2}\hat{a}_{2}}e^{\mu _{1}\hat{a}_{1}^{\dag }}e^{\mu _{2}\hat{a}%
_{2}^{\dag }}\rho _{in}e^{\nu _{1}\hat{a}_{1}}e^{\nu _{2}\hat{a}_{2}}  \notag
\\
& e^{n_{1}\hat{a}_{1}^{\dag }}e^{n_{2}\hat{a}_{2}^{\dag }}\left\vert
0\right\rangle _{1}\left\vert 0\right\rangle _{2}|_{\mu _{1}=\mu
_{2}=v_{1}=\nu _{2}=m_{1}=m_{2}=n_{1}=n_{2}=0}.  \tag{C3}
\end{align}

(I) For case CS-CS, we have
\end{subequations}
\begin{subequations}
\begin{align}
& p_{h_{1},h_{2},t_{1},t_{2},k_{1}k_{2},l_{1}l_{2}}^{cc}  \notag \\
& =\frac{1}{\sqrt{k_{1}!l_{1}!k_{2}!l_{2}!}}\partial
_{m_{1}}^{k_{1}}\partial _{m_{2}}^{k_{2}}\partial _{n_{1}}^{l_{1}}\partial
_{n_{2}}^{l_{2}}\partial _{\mu _{1}}^{h_{1}}\partial _{\mu
_{2}}^{h_{2}}\partial _{\nu _{1}}^{t_{1}}\partial _{\nu _{2}}^{t_{2}}  \notag
\\
& e^{-\left\vert z_{1}\right\vert ^{2}+m_{1}\left( \mu _{1}+z_{1}\right)
+n_{1}\left( \nu _{1}+z_{1}^{\ast }\right) -\left\vert z_{2}\right\vert
^{2}+m_{2}\left( \mu _{2}+z_{2}\right) +n_{2}\left( \nu _{2}+z_{2}^{\ast
}\right) }  \notag \\
& |_{m_{1}=\mu _{1}=v_{1}=n_{1}=m_{2}=\mu _{2}=\nu _{2}=n_{2}=0}.  \tag{C4}
\end{align}

(II) For case TS-TS, we have
\end{subequations}
\begin{subequations}
\begin{align}
& p_{h_{1},h_{2},t_{1},t_{2},k_{1},k_{2},l_{1},l_{2}}^{tt}  \notag \\
& =\frac{A^{-1}}{\sqrt{k_{1}!l_{1}!k_{2}!l_{2}!}}\partial
_{m_{1}}^{k_{1}}\partial _{m_{2}}^{k_{2}}\partial _{n_{1}}^{l_{1}}\partial
_{n_{2}}^{l_{2}}\partial _{\mu _{1}}^{h_{1}}\partial _{\mu
_{2}}^{h_{2}}\partial _{\nu _{1}}^{t_{1}}\partial _{\nu _{2}}^{t_{2}}  \notag
\\
& e^{m_{1}\mu _{1}+m_{2}\mu _{2}+\nu _{1}n_{1}+\nu _{2}n_{2}+\frac{\bar{n}%
_{1}}{\bar{n}_{1}+1}m_{1}n_{1}+\frac{\bar{n}_{2}}{\bar{n}_{2}+1}m_{2}n_{2}}
\notag \\
& |_{m_{1}=\mu _{1}=v_{1}=n_{1}=m_{2}=\mu _{2}=\nu _{2}=n_{2}=0}.  \tag{C5}
\end{align}

(III) For case SV-SV, we have
\end{subequations}
\begin{subequations}
\begin{align}
& p_{h_{1},h_{2},t_{1},t_{2},k_{1}k_{2},l_{1}l_{2}}^{ss}  \notag \\
& =\frac{\left( \kappa _{1}\kappa _{2}\right) ^{-1/2}}{\sqrt{%
k_{1}!l_{1}!k_{2}!l_{2}!}}\partial _{m_{1}}^{k_{1}}\partial
_{m_{2}}^{k_{2}}\partial _{n_{1}}^{l_{1}}\partial _{n_{2}}^{l_{2}}\partial
_{\mu _{1}}^{h_{1}}\partial _{\mu _{2}}^{h_{2}}\partial _{\nu
_{1}}^{t_{1}}\partial _{\nu _{2}}^{t_{2}}  \notag \\
& e^{m_{1}\mu _{1}+\frac{\lambda _{1}}{2}m_{1}^{2}+\nu _{1}n_{1}+\frac{%
\lambda _{1}}{2}n_{1}^{2}+m_{2}\mu _{2}+\frac{\lambda _{2}}{2}m_{2}^{2}+\nu
_{2}n_{2}+\frac{\lambda _{2}}{2}n_{2}^{2}}  \notag \\
& |_{\mu _{1}=\mu _{2}=\nu _{1}=\nu _{2}=m_{1}=m_{2}=n_{1}=n_{2}=0}.
\tag{C6}
\end{align}

(IV) For case TMSV, we have

\end{subequations}
\begin{subequations}
\begin{align}
& p_{h_{1},h_{2},t_{1},t_{2},k_{1}k_{2},l_{1}l_{2}}^{tms}  \notag \\
& =\frac{\kappa ^{-1}}{\sqrt{k_{1}!k_{2}!l_{1}!l_{2}!}}\partial
_{m_{1}}^{k_{1}}\partial _{m_{2}}^{k_{2}}\partial _{n_{1}}^{l_{1}}\partial
_{n_{2}}^{l_{2}}\partial _{\mu _{1}}^{h_{1}}\partial _{\mu
_{2}}^{h_{2}}\partial _{\nu _{1}}^{t_{1}}\partial _{\nu _{2}}^{t_{2}}  \notag
\\
& e^{m_{1}\mu _{1}+m_{2}\mu _{2}+n_{1}\nu _{1}+n_{2}\nu _{2}+\lambda
m_{1}m_{2}+\lambda n_{1}n_{2}}  \notag \\
& |_{m_{1}=m_{2}=\mu _{1}=\mu _{2}=v_{1}=\nu _{2}=n_{1}=n_{2}=0}.  \tag{C7}
\end{align}

\subsection*{Appendix D: $X^{T_{2}}$s of these states and their eigenvalues}

Using $p_{k_{1},k_{2},l_{1},l_{2}}$ in Appendix C, we obtain $X^{T_{2}}$s
and calculate their eigenvalues.

(I) Case CS-CS: For $\rho ^{cc}(z_{1},z_{2})$, we obtain matrix
\end{subequations}
\begin{subequations}
\begin{equation}
X_{\rho ^{cc}(z_{1},z_{2})}^{T_{2}}=\frac{1}{M}\left(
\begin{array}{cccc}
1 & z_{2} & z_{1}^{\ast } & z_{2}z_{1}^{\ast } \\
z_{2}^{\ast } & \left\vert z_{2}\right\vert ^{2} & z_{1}^{\ast }z_{2}^{\ast }
& \left\vert z_{2}\right\vert ^{2}z_{1}^{\ast } \\
z_{1} & z_{1}z_{2} & \left\vert z_{1}\right\vert ^{2} & z_{2}\left\vert
z_{1}\right\vert ^{2} \\
z_{1}z_{2}^{\ast } & z_{1}\left\vert z_{2}\right\vert ^{2} & z_{2}^{\ast
}\left\vert z_{1}\right\vert ^{2} & \left\vert z_{1}\right\vert
^{2}\left\vert z_{2}\right\vert ^{2}%
\end{array}%
\right)   \tag{D1}
\end{equation}%
with $M=(1+\left\vert z_{1}\right\vert ^{2})(1+\left\vert z_{2}\right\vert
^{2})$, the eigenvalues of which include $1$, $0$, $0$, and $0$. Obviously,
there is no negative eigenvalue. For $\rho _{cc,\varphi }(z_{1},z_{2})$, we
obtain matrix
\end{subequations}
\begin{subequations}
\begin{equation}
X_{\rho _{cc,\varphi }(z_{1},z_{2})}^{T_{2}}=\left(
\begin{array}{cccc}
0 & 0 & 0 & \frac{e^{-i\varphi }}{N_{cc}} \\
0 & \frac{1}{N_{cc}} & 0 & \frac{z_{1}+e^{-i\varphi }z_{2}}{N_{cc}} \\
0 & 0 & \frac{1}{N_{cc}} & \frac{e^{-i\varphi }z_{1}^{\ast }+z_{2}^{\ast }}{%
N_{cc}} \\
\frac{e^{i\varphi }}{N_{cc}} & \frac{z_{1}^{\ast }+e^{i\varphi }z_{2}^{\ast }%
}{N_{cc}} & \frac{e^{i\varphi }z_{1}+z_{2}}{N_{cc}} & \frac{\left\vert
z_{1}+e^{-i\varphi }z_{2}\right\vert ^{2}}{N_{cc}}%
\end{array}%
\right)   \tag{D2}
\end{equation}%
with $N_{cc}=2+\left\vert z_{1}+e^{-i\varphi }z_{2}\right\vert ^{2}$, the
eigenvalues of which include $-N_{cc}^{-1}$, $N_{cc}^{-1}$, $\frac{1}{2}(1-%
\sqrt{1-4N_{cc}^{-2}})$, and $\frac{1}{2}(1-\sqrt{1+4N_{cc}^{-2}})$. It has
a negative eigenvalue $-N_{cc}^{-1}$.

(II) Case TS-TS: For $\rho ^{tt}(\overline{n}_{1},\overline{n}_{2})$, we
obtain matrix
\end{subequations}
\begin{subequations}
\begin{equation}
X_{\rho ^{tt}(\overline{n}_{1},\overline{n}_{2})}^{T_{2}}=\left(
\begin{array}{cccc}
\frac{A}{B} & 0 & 0 & 0 \\
0 & \frac{\left( 1+\overline{n}_{1}\right) \overline{n}_{2}}{B} & 0 & 0 \\
0 & 0 & \frac{\overline{n}_{1}\left( 1+\overline{n}_{2}\right) }{B} & 0 \\
0 & 0 & 0 & \frac{\overline{n}_{1}\overline{n}_{2}}{B}%
\end{array}%
\right)   \tag{D3}
\end{equation}%
with $A=\left( 1+\overline{n}_{1}\right) \left( 1+\overline{n}_{2}\right) $
and $B=\left( 1+2\overline{n}_{1}\right) \left( 1+2\overline{n}_{2}\right) $%
, the eigenvalues of which include $\overline{n}_{1}\overline{n}_{2}B^{-1}$,
$AB^{-1}$, $\overline{n}_{1}\left( 1+\overline{n}_{2}\right) B^{-1}$, and $%
\allowbreak \left( 1+\overline{n}_{1}\right) \overline{n}_{2}B^{-1}$.
Clearly, there was no negative eigenvalue. For $\rho _{tt,\varphi }(%
\overline{n}_{1},\overline{n}_{2})$, we obtain matrix
\end{subequations}
\begin{subequations}
\begin{equation}
X_{\rho _{tt,\varphi }(\overline{n}_{1},\overline{n}_{2})}^{T_{2}}=\left(
\begin{array}{cccc}
0 & 0 & 0 & \frac{Ae^{i\varphi }}{2A+\Gamma } \\
0 & \frac{A}{2A+\Gamma } & 0 & 0 \\
0 & 0 & \frac{A}{2A+\Gamma } & 0 \\
\frac{Ae^{-i\varphi }}{2A+\Gamma } & 0 & 0 & \frac{\Gamma }{2A+\Gamma }%
\end{array}%
\right) ,  \tag{D4}
\end{equation}%
the eigenvalues of which include $A/\left( 2A+\Gamma \right) $, $A/\left(
2A+\Gamma \right) $, $-\frac{1}{2}(\sqrt{4A^{2}+\Gamma ^{2}}-\Gamma )/\left(
2A+\Gamma \right) $, and $\frac{1}{2}(\sqrt{4A^{2}+\Gamma ^{2}}+\Gamma
)/\left( 2A+\Gamma \right) $ with $\Gamma =\overline{n}_{1}+\overline{n}%
_{2}+2\overline{n}_{1}\overline{n}_{2}$. It has a negative eigenvalue.

(III) Case SV-SV: For $\rho ^{ss}(r_{1},r_{2})$, we obtain matrix
\end{subequations}
\begin{subequations}
\begin{equation}
X_{\rho ^{ss}(r_{1},r_{2})}^{T_{2}}=\left(
\begin{array}{cccc}
1 & 0 & 0 & 0 \\
0 & 0 & 0 & 0 \\
0 & 0 & 0 & 0 \\
0 & 0 & 0 & 0%
\end{array}%
\right) ,  \tag{D5}
\end{equation}%
the eigenvalues of which include $1$, $0$, $0$, and $0$. Obviously, there is
no negative eigenvalue. For $\rho _{ss,\varphi }(r_{1},r_{2})$, we obtain
matrix
\end{subequations}
\begin{subequations}
\begin{equation}
X_{\rho _{ss,\varphi }(r_{1},r_{2})}^{T_{2}}=\left(
\begin{array}{cccc}
0 & 0 & 0 & \frac{e^{i\varphi }}{2} \\
0 & \frac{1}{2} & 0 & 0 \\
0 & 0 & \frac{1}{2} & 0 \\
\frac{e^{-i\varphi }}{2} & 0 & 0 & 0%
\end{array}%
\right) ,  \tag{D6}
\end{equation}%
the eigenvalues of which include $-\frac{1}{2}$, $\frac{1}{2}$, $\frac{1}{2}$%
, and$\frac{1}{2}$. It has one negative eigenvalue $-\frac{1}{2}$.

(IV) Case TMSV: For $\rho ^{tms}(r)$, we obtain matrix

\end{subequations}
\begin{subequations}
\begin{equation}
X_{\rho ^{tms}(r)}^{T_{2}}=\left(
\begin{array}{cccc}
\frac{1}{1+\lambda ^{2}} & 0 & 0 & 0 \\
0 & 0 & \frac{\lambda }{1+\lambda ^{2}} & 0 \\
0 & \frac{\lambda }{1+\lambda ^{2}} & 0 & 0 \\
0 & 0 & 0 & \frac{\lambda ^{2}}{1+\lambda ^{2}}%
\end{array}%
\right)   \tag{D7}
\end{equation}%
the eigenvalues of which include $\frac{1}{\lambda ^{2}+1}$, $-\frac{\lambda
}{\lambda ^{2}+1}$, $\frac{\lambda ^{2}}{\lambda ^{2}+1}$, and $\frac{%
\lambda }{\lambda ^{2}+1}$. Obviously, there is one negative eigenvalue $-%
\frac{\lambda }{\lambda ^{2}+1}$. For $\rho _{ss,\varphi }(r_{1},r_{2})$, we
obtain matrix
\end{subequations}
\begin{subequations}
\begin{equation}
X_{\rho _{tms,\varphi }(r)}^{T_{2}}=\left(
\begin{array}{cccc}
0 & 0 & 0 & \frac{e^{i\varphi }}{2} \\
0 & \frac{1}{2} & 0 & 0 \\
0 & 0 & \frac{1}{2} & 0 \\
\frac{e^{-i\varphi }}{2} & 0 & 0 & 0%
\end{array}%
\right) ,  \tag{D8}
\end{equation}%
the eigenvalues of which include $-\frac{1}{2}$, $\frac{1}{2}$, $\frac{1}{2}$%
, and $\frac{1}{2}$. It has one negative eigenvalue $-\frac{1}{2}$.

\subsection*{Appendix E: Wigner functions of these states}

In normal ordering form, we have $\Pi \left( \beta _{j}\right) =\frac{2}{\pi
}\colon e^{-2(\hat{a}_{j}^{\dag }-\beta _{j}^{\ast })(\hat{a}_{j}-\beta
_{j})}\colon $ ( $\colon \cdots \colon $\ denotes normal ordering). As long
as we know $W_{\rho _{h_{1},h_{2},t_{1},t_{2}}}\left( \beta _{1},\beta
_{2}\right) $, we can obtain $W_{\rho _{in}}\left( \beta _{1},\beta
_{2}\right) =W_{\rho _{0,0,0,0}}\left( \beta _{1},\beta _{2}\right) $ and
\end{subequations}
\begin{subequations}
\begin{align}
& W_{\rho _{out,\varphi }}\left( \beta _{1},\beta _{2}\right)   \notag \\
& =N^{-1}[W_{\rho _{1,0,1,0}}\left( \beta _{1},\beta _{2}\right) +W_{\rho
_{0,1,0,1}}\left( \beta _{1},\beta _{2}\right)   \notag \\
& +e^{-i\varphi }W_{\rho _{1,0,0,1}}\left( \beta _{1},\beta _{2}\right)
+e^{i\varphi }W_{\rho _{0,1,1,0}}\left( \beta _{1},\beta _{2}\right) ].
\tag{E1}
\end{align}%
Next we give $W_{\rho _{h_{1},h_{2},t_{1},t_{2}}}\left( \beta _{1},\beta
_{2}\right) $\ for each case.

(I) For case CS-CS, we have
\end{subequations}
\begin{subequations}
\begin{align}
& W_{\rho _{h_{1},h_{2},t_{1},t_{2}}^{cc}}\left( \beta _{1},\beta _{2}\right)
\notag \\
& =W_{\rho ^{cc}(z_{1},z_{2})}\left( \beta _{1},\beta _{2}\right) \partial
_{\mu _{1}}^{h_{1}}\partial _{\mu _{2}}^{h_{2}}\partial _{\nu
_{1}}^{t_{1}}\partial _{\nu _{2}}^{t_{2}}  \notag \\
& e^{\left( 2\beta _{1}^{\ast }-\allowbreak z_{1}^{\ast }\right) \mu
_{1}+\left( 2\beta _{1}-z_{1}\right) \nu _{1}-\mu _{1}\nu _{1}}  \notag \\
& e^{\left( 2\beta _{2}^{\ast }-z_{2}^{\ast }\right) \mu _{2}+\left( 2\beta
_{2}-z_{2}\right) \nu _{2}-\mu _{2}\nu _{2}}|_{\mu _{1}=\mu _{2}=v_{1}=\nu
_{2}=0}.  \tag{E2}
\end{align}

(II) For case TS-TS, we have
\end{subequations}
\begin{subequations}
\begin{align}
& W_{\rho _{h_{1},h_{2},t_{1},t_{2}}^{tt}}\left( \beta _{1},\beta _{2}\right)
\notag \\
& =W_{\rho ^{tt}(\bar{n}_{1},\bar{n}_{2})}\left( \beta _{1},\beta
_{2}\right) \partial _{\mu _{1}}^{h_{1}}\partial _{\mu _{2}}^{h_{2}}\partial
_{\nu _{1}}^{t_{1}}\partial _{\nu _{2}}^{t_{2}}  \notag \\
& e^{\epsilon _{1}(2\beta _{1}^{\ast }\mu _{1}+2\beta _{1}\nu _{1}-\mu
_{1}\nu _{1})+\epsilon _{2}(2\beta _{2}^{\ast }\mu _{2}+2\beta _{2}\nu
_{2}-\mu _{2}\nu _{2})}  \notag \\
& |_{\mu _{1}=\mu _{2}=v_{1}=\nu _{2}=0}.  \tag{E3}
\end{align}

(III) For case SV-SV, we have
\end{subequations}
\begin{subequations}
\begin{align}
& W_{\rho _{h_{1},h_{2},t_{1},t_{2}}^{ss}}\left( \beta _{1},\beta _{2}\right)
\notag \\
& =W_{\rho ^{ss}(r_{1},r_{2})}\left( \beta _{1},\beta _{2}\right) \partial
_{\mu _{1}}^{h_{1}}\partial _{\mu _{2}}^{h_{2}}\partial _{\nu
_{1}}^{t_{1}}\partial _{\nu _{2}}^{t_{2}}  \notag \\
& e^{\kappa _{1}(2\allowbreak \beta _{1}^{\ast }\mu _{1}+2\beta _{1}\nu
_{1}-\nu _{1}\mu _{1})+\kappa _{2}(2\beta _{2}^{\ast }\mu _{2}+2\beta
_{2}\nu _{2}-\nu _{2}\mu _{2})}  \notag \\
& e^{\lambda _{1}\kappa _{1}(+\allowbreak \frac{1}{2}\nu _{1}^{2}+\frac{1}{2}%
\mu _{1}^{2}-2\nu _{1}\beta _{1}^{\ast }\allowbreak -2\beta _{1}\mu _{1})}
\notag \\
& e^{\lambda _{2}\kappa _{2}(+\allowbreak \frac{1}{2}\nu _{2}^{2}+\frac{1}{2}%
\mu _{2}^{2}-2\nu _{2}\beta _{2}^{\ast }\allowbreak -2\beta _{2}\mu _{2})}
\notag \\
& |_{\mu _{1}=\mu _{2}=v_{1}=\nu _{2}=0}.  \tag{E4}
\end{align}

(IV) For case TMSV, we have
\end{subequations}
\begin{subequations}
\begin{align}
& W_{\rho _{h_{1},h_{2},t_{1},t_{2}}^{tms}}\left( \beta _{1},\beta
_{2}\right)  \notag \\
& =W_{\rho ^{tms}(r)}\left( \beta _{1},\beta _{2}\right) \partial _{\mu
_{1}}^{h_{1}}\partial _{\mu _{2}}^{h_{2}}\partial _{\nu
_{1}}^{t_{1}}\partial _{\nu _{2}}^{t_{2}}  \notag \\
& e^{\kappa \left( 2\allowbreak \beta _{1}^{\ast }\mu _{1}+2\beta _{1}\nu
_{1}+2\beta _{2}^{\ast }\mu _{2}+2\beta _{2}\nu _{2}-\nu _{1}\mu _{1}-\nu
_{2}\mu _{2}\right) }  \notag \\
& e^{\lambda \kappa \left( \nu _{1}\nu _{2}+\mu _{1}\mu _{2}-2\beta _{1}\mu
_{2}-2\beta _{1}^{\ast }\nu _{2}-2\beta _{2}^{\ast }\nu _{1}\allowbreak
-2\beta _{2}\mu _{1}\right) }  \notag \\
& |_{\mu _{1}=\mu _{2}=v_{1}=\nu _{2}=0}.  \tag{E5}
\end{align}

\begin{acknowledgments}
This paper was supported by the National Natural Science Foundation of China
(Grant No. 11665013). We would like to thank Qiongyi He, Yu Xiang, Shengli
Zhang, Bixuan Fan and Zhenglu Duan for useful discussions.
\end{acknowledgments}

\end{subequations}

\end{document}